# Evidence from EXAFS for Different Ta/Ti Site Occupancy in High Critical Current Density $Nb_3Sn$ Superconductor Wires

Steve M. Heald,[1] Chiara Tarantini,[2*] Peter J. Lee,[2] Michael Brown,[2] ZuHawn Sung,[2] Arup K. Ghosh,[3] David C. Larbalestier[2]

[1] *Advanced Photon Source, Argonne National Lab, Argonne IL 60439 USA*

[2] *National High Magnetic Field Laboratory, Florida State University, Tallahassee, Florida 32310, USA*

[3] *Retired, formerly Magnet Division, Brookhaven National Laboratory, Upton, New York 11973, USA*

To meet critical current density, $J_c$, targets for the Future Circular Collider (FCC), the planned replacement for the Large Hadron Collider (LHC), the high field performance of $Nb_3Sn$ must be improved, but champion $J_c$ values have remained static for the last 10 years. Making the A15 phase stoichiometric and enhancing the upper critical field $H_{c2}$ by Ti or Ta dopants are the standard strategies for enhancing high field performance but detailed recent studies show that even the best modern wires have broad composition ranges. To assess whether further improvement might be possible, we employed EXAFS to determine the lattice site location of dopants in modern high-performance $Nb_3Sn$ strands with $J_c$ values amongst the best so far achieved. Although Ti and Ta primarily occupy the Nb sites in the A15 structure, we also find significant Ta occupancy on the Sn site. These findings indicate that the best performing Ti-doped stand is strongly sub-stoichiometric in Sn and that antisite disorder likely explains its high average $H_{c2}$ behavior. These new results suggest an important role for dopant and antisite disorder in minimizing superconducting property distributions and maximizing high field $J_c$ properties.

[*]e-mail: tarantini@asc.magnet.fsu.edu

**Introduction**

Nb$_3$Sn superconducting wires are the first choice for magnet applications at fields above 10-11 T, such as NMR spectroscopy, compact cyclotrons, magnetically confined fusion reactors and the next generation of particle accelerators because of their relatively low cost compared to HTS superconductors. In particular, Nb$_3$Sn wires are being used to wind magnets for the Hi-Lumi upgrade of the Large Hadron Collider[1-3] and they are the most technologically ready candidates for the LHC replacement, the envisioned Future Circular Collider (FCC)[4]. However, the performance of these wires must be improved to meet the FCC requirements[5], so there is an important need to understand the limitations of the best currently produced wires. One area of uncertainty is how close we are to full optimization of the wires because Nb$_3$Sn is in fact a compound with a range of Sn composition and the critical temperature, $T_c$, upper critical field, $H_{c2}$, and critical current density, $J_c$, are all sensitive to the composition, which is never without variation within real wires. In binary Nb$_3$Sn, $T_c$ and $H_{c2}$ increase as the ratio of Nb:Sn approaches the stoichiometric value of 3.[6] However, dopants (typically Ti and/or Ta) are required for high field operation. In fact these dopants can raise the $H_{c2}(0)$ by ~3 T [7] (conventionally attributed to additional electron scattering increasing the normal state resistivity and thus $H_{c2}(0)$)[8]. For the strand design closest to meeting the FCC critical current target (which uses the "internal-Sn" process), we have previously shown that doping with Ti rather than Ta produces a more homogeneous A15 composition and a tighter distribution of $T_c$. Furthermore, Ti appears to accelerate the diffusion reaction that creates the A15 phase from its Sn and Nb constituents[9]. An early EXAFS (Extended X-ray Absorption Fine Structure) study of Ta doped samples [10] and an ALCHEMI (Atom Location by Channeling Enhanced Microanalysis) study of both Ti-doped and Ta-doped samples [11] using older generation strands indicated that both dopants substitute on the Nb sites. However, more recent studies hypothesized that Ti might substitute on the Sn site[12,13]. This hypothesis seemed to better fit the normal and superconducting properties of Ti-doped rather than Ta-doped Nb$_3$Sn and their average Sn chemical compositions. In ref. 9 we found that the Ti-doped wire had very high $J_c$ values (3035 A/mm$^2$ at 12 T and 4.2 K) and a relatively narrow specific heat $T_c$-distribution at 15 T despite measuring only 23.1 at%Sn in the A15 layer. Ti doping on the Nb site would lead to a strongly off-stoichiometric A15, whereas the Flukiger et al.[12,13] suggestion of Ti being on the Sn site would lead to a composition much closer to stoichiometry. This apparent inconsistency between studies persuaded us to revisit the early EXAFS study with our modern high-$J_c$ wires with Ta, Ti and Ta+Ti-doping.

The extended x-ray absorption fine structure (EXAFS) is sensitive to the local environment of the dopant atoms. It occurs at x-ray energies above the absorption edge energy of the element under study. Thus, by tuning the x-ray energy to different absorption edges, the local structure of individual elements in a complex material can be determined. In general terms, considering only the dominant single scattering paths, the EXAFS function, $\chi(k)$, can be expressed as:

$$\chi(k) = \frac{\mu(k) - \mu_0(k)}{\mu_0(k)} = \sum_j \frac{N_j}{kR_j^2} A_j(k) \sin[2kR_j + \psi_j(k)], \qquad k = \frac{\sqrt{2m_e(E - E_0)}}{\hbar}.$$

where $\mu$ is the x-ray absorption coefficient, $\mu_0$ is the smoothly varying absorption of an isolated atom, $E_0$ is the absorption edge energy of the element under study, and $k$ is the wave vector of the photoelectron that is ejected when the absorption edge energy is crossed. The sum is over the local coordination shells of atoms, where $N$ is the number of atoms in the shell, $R$ is the distance from the absorbing atom, $\psi(k)$ is a phase shift term, and $A(k)$ is an overall amplitude factor that includes a Debye-Waller like factor, $e^{-2k^2\sigma^2}$, where $\sigma^2$ is a measure of the disorder over the distance $R$. The disorder can be structural due to a rearrangement of the atoms around the dopant or vibrational due to the natural vibrations of the atoms about their lattice sites. Because each shell of atoms give rise to a distinct frequency, $2kR+\psi$, a Fourier transform of the spectrum can be used to separate the contributions of the different shells.

For the $Nb_3Sn$ structure there is a distinct difference in the local environment of the two sites. The Nb atoms have three closely spaced nearby coordination shells, while the Sn site has a single nearest neighbor shell (Figure 1). For a reference sample, we characterized a bulk binary $Nb_3Sn$ sample fabricated from just Nb and Sn powders. Table I summarizes the parameters for the two sites. Because of the difference in the coordination shells, the EXAFS from the Nb and Sn sites is distinctly different. Figure 1 shows the Fourier transformed spectra from Nb and Sn in pure $Nb_3Sn$. The three closely spaced nearest neighbor shells for the Nb site gives rise to a distinct 3-peak structure in the Fourier transform. These can be well fit using the parameters given in Table I. In this fitting the number of neighbors N was fixed, and theoretical amplitudes and phase shifts were calculated using FEFF 7 [14]. The fitted distances agree well with the crystallographic values,[15] and the Debye-Waller factors are reasonable. Also, note that the data from the two edges were fit independently. The R and $\sigma^2$ values for the Nb-Sn and Sn-Nb paths should be the same. Although unconstrained, the close agreement between these two results supports the accuracy of the fitting model.

**Results**

*Samples and initial characterizations*

A binary $Nb_3Sn$ bulk sample to serve as a standard was prepared with a nominal 25 at% Sn (see Methods) and the composition was verified by EDS (Energy Dispersive Spectroscopy) to be between 24.3 and 25.3 at.% Sn. The descriptions and basic properties of the wires under investigation are summarized in Table II. To compare Ti, Ti + Ta and Ta doping, we chose a first series of samples with nominally identical designs but with doping changes only, as detailed in Figure 2. These strands were of the internal-Sn design ($Nb_3Sn$ is formed by diffusion of Sn from central Sn cores to surrounding Nb filaments through a Cu matrix) manufactured by the Restacked Rod Process (RRP®) by Oxford Superconducting Technologies (now Bruker-OST). The 54 internal-Sn sub-elements were hexagonally stacked around 7 Cu blanks inside an outer Cu tube to create the "54/61" multifilamentary composites. In the Ti-doped strand (OST billet #9415-BE), ~2 at% Ti was introduced by inserting 30 Nb-47wt.%Ti rods into the filament packs within each sub-element. This sample is referred to here as Ti#1. For the Ta-doped strand (OST #8781) the filaments and the diffusion barriers were 4 at% Ta instead of pure Nb [10]. The Ta+Ti composite (OST #9362-5) used a combination of the 4 at.% Ta alloy filaments and barrier with 16 Nb-Ti filaments produced a composition of ~1 at.% Ti + 4 at.% Ta. The samples were part of an extensive strain study [16] using a common reaction heat treatment (HT) of 40 h at 640 °C. They were slightly underreacted to avoid any Sn diffusion into the stabilizer Cu. In addition, a second similarly made RRP® "108/127" wire series was included: a Ti-doped (Ti#2) and a Ta-doped (Ta#2) wire from our earlier $T_c$-distribution comparison study [9] (OST #14895 and #12879). These samples were heat treated by Arup Ghosh at Brookhaven National Laboratory at 662-665 °C for 48 h and reacted close to their maximum $J_c$ as is reflected in the ≈3000 A/mm$^2$ (12 T, 4.2 K) performance. Because of the different heat treatments, the two series show differences in both $J_c$ and the extrapolated Kramer field ($J_c$ ~ 2620-2870 A/mm$^2$, $\mu_0 H_k$ ~ 22.7-24.6 T for the 54/61 series and $J_c$ ~3130-3210 A/mm$^2$, $\mu_0 H_k$ ~ 23.8-25.5 T for the 108/127 series). Noticeable is also the larger $H_k$ of the Ta+Ti strand, despite its inferior $J_c$(12T) performance when compared with both Ti#1 and Ta#1. For completeness the "54/61" series was analyzed by specific heat to determine the 0-15 T $T_c$-distributions, in the same way that we previously characterized the "108/127" wires[9]. This characterization is shown in Figure 3. Besides confirming the results found for the "108/127" series wires (better homogeneity of the Ti-doped sample with respect to the Ta-doped one), reveals that the Ta+Ti strand has a tighter in-field $T_c$-distribution than both Ta- or Ti-doped samples, despite a quite broad zero-field $T_c$-distribution. The WHH extrapolations[17] of $H_{c2}(0)$ obtained for all the samples from the $T_c$-distributions are reported in Table II as well. Similarly to $H_k$, also for $H_{c2}(0)$ we found a higher value in the double-doped Ta+Ti wire than with Ta#1 and Ti#1.

### *EXAFS characterization*

Most of the EXAFS measurements were made using the 20-ID-B microprobe station at the Advanced Photon Source[18]. This provided a 3 µm beam that could be focused onto the narrow $Nb_3Sn$ regions in the strands. Figure 2 shows some examples of fluorescence maps from one of the samples. These were used to select the proper area for the EXAFS measurements (since the core is not fully dense and the incident/emission beams to obtain the maps are at +/- 45° from the surface normal, the apparent high Ti (Sn) composition on the right side of the sub-element is an artifact due to the experiment geometry). The EXAFS set-up is described in the Methods. Supplementary Figures S1 and S2 show the raw χ(k) data measured on the Ta and Ti edges for all samples. The corresponding Fourier transformed data are plotted in Figure 4 and Figure 5. Also shown in these figures are fits to the data limited to the main structure region using the Artemis software[19], which allows the dopants to enter either site. Because the fitting range is limited, it was not possible to vary all of the parameters corresponding to the 4 separate scattering paths of the two sites (see Figure 1). To minimize the variable parameters, there were two main assumptions. First the coordination numbers were assumed to remain the same. The second is the assumption that the distance and Debye Waller factors for the second shell of the Nb site are the same as for the first shell of the Sn site. While strictly correct for the binary sample, it is, however, a reasonable assumption also for the doped compounds. The overall amplitude factor was allowed to vary. It ended up in the range of 0.65 to 0.72 for the three samples containing Ti and about 1.1-1.3 for the Ta samples.

The fitting results for the Ta data are shown in Figure 4 and the parameters summarized in Table III. From the transforms, it is obvious that most of the Ta is in the Nb sites (3-peak structure). However, the best fits were obtained with significant Ta occupation of Sn sites (solid green lines): about 30% for Ta#1, 32% for the Ta+Ti and 21% for Ta#2. For comparison these data were also fit with Ta on the Nb site only. These fits are also shown in Figure 4 (dashed red lines) where it is clear that the single-site fits are inferior.

For the Ti samples, the fits are shown in Figure 5 and summarized in Table IV. Again, the transforms indicate that the Nb site is preferred but, in this case, the Ti occupancy of the Sn site in the two-site fit always refined to 0. Thus, for the Ti doped samples, the EXAFS indicates that Ti sits only on the Nb sites.

### *Analysis of disorder and correlation to $H_{c2}$*

At the root of our concern for the site occupancy is consideration of whether there might be an effect on $H_{c2}$, epscially if site disorder could add to the electron scattering that drives up $H_{c2}$. Indeed the EXAFS results provide us information useful to correlate the disorder introduced into the A15 phase with the $H_{c2}$ trend, taking into account the WHH formula $H_{c2}(0) = 0.69 T_c \, dH_{c2}/dT|_{T_c}$ with $dH_{c2}/dT|_{T_c} \propto \gamma \rho_0$ (where $\gamma$ and $\rho_0$ are the Sommerfeld constant and the normal state resistivity)[17]. Since $\gamma$ is proportional to the density of states $N_F$ and $\rho_0$ is proportional to $1/N_F \tau$ (with $\tau$ being the scattering rate), $H_{c2}$ can be written as $H_{c2}(0) \propto T_c/\tau$. Moreover, the scattering rate is related to the long-range order (LRO) parameter $\eta$ by the relation $1/\tau \propto (1-\eta^2)$,[20] so $H_{c2}$ should follow the relation $H_{c2}(0) \propto T_c(1-\eta^2)$. Before estimating the LRO parameter, we need to clarify what type of disorder is present in the A15 phase. In fact, Table V shows that the Nb/Sn ratio estimated by EDS in the A15 layer is off-stoichiometric in all samples. If the dopant occupancy $x$ on the Sn site obtained by EXAFS is taken into account, the resulting ratio (Nb+Ti+(1-$x$)Ta)/(Sn+$x$Ta) is in most cases even more strongly off-stoichiometry: this is particularly evident in the Ti-doped samples where it exceeds 3.4. This off-stoichiometry implies that the dopants are not the only source of disorder in the A15 structure but either vacancy (i.e. empty site) or antisite (i.e. atom occupying the "wrong" site) disorder has to be considered. By *ab initio* calculations, Besson *et al.*[21] estimated the defect structure in Nb$_3$Sn finding that it is of antisite nature: in fact, their results indicate that the fraction of defects by vacancies is at least 7 orders of magnitude smaller than by antisite disorder at 1000 K (more than 20 orders of magnitude at 300 K). This means that vacancies are extremely rare in Nb$_3$Sn and that off-stoichiometry occurs by antisite disorder.

For an alloy with two sublattices (as in the case of the A15 structure, $A_3B$) two distinct LRO parameters have to be defined as: $\eta_A = c_A(A) - c_A(B)$ and $\eta_B = c_B(B) - c_B(A)$ where $c_\kappa(\alpha)$ is the fraction of the $\alpha$ sites occupied by the $\kappa$ element[22]. The LRO parameter is then defined by the fractional difference between elements siting on the "right" sites and those sitting on the "wrong" sites and in general it varies between 1 (perfectly ordered system) and 0 (a completely disordered system with random site occupancy). With ternary additions (X), as in our case, the LRO is always less than 1, requiring use of the parameters $c_X(A)$ and $c_X(B)$. To calculate the LRO parameters $\eta_{Nb}$ and $\eta_{Sn}$ for our samples we can take into account the global composition determined by EDS (Table II), the occupancy resulting from the EXAFS (Table III and IV) and we can impose the 3:1 ratio of the $A_3B$ structure as being fulfilled by antisite substitutions (following the conclusions of the ref. 21 calculations). It is important to notice that this evaluation will underestimate the disorder because disorder induced by antisite exchange (i.e. Nb and Sn atoms exchanging sites) is unknown. Although this contribution might be limited for single doping, it is likely important for the Ta+Ti sample (because Ta and Ti favor opposite antisite disorder in the 54/61 series). In this case, we include a small amount of exchange antisites proportional to the amount of antisites that Ti generates in the Ti-doped samples reduced by the amount of Ta on the Sn sites (Ta on Sn already acts like an exchange antisite). The resulting values are reported in Table V (the values in parentheses for the Ta+Ti sample do not include any exchange antisites and they are shown for comparison: clearly, they are incompatible with the high $H_{c2}$ of Ta+Ti). In both series, Ta-doped samples are the more ordered (highest $\eta_{Nb}$ and $\eta_{Sn}$ values) whereas Ti-doped samples are more disordered than Ta-doped ones. We also found that Ta+Ti sample from the 54/61 series is more disordered than either Ta#1 or the Ti#1.

Due to the complexity of the system (ternary/quaternary phase in a structure with 2 sublattices), in order to test the presence of the $H_{c2}$-disorder relationship given by $H_{c2}(0) \propto T_c(1-\eta^2)$, we used an effective $\eta$ value calculated as the weighted average of $\eta_{Nb}$ and $\eta_{Sn}$. This trend is shown in Figure 6. Despite the approximations of the disorder estimate, four of the samples follow a linear trend and only the least disordered sample (Ta#1) appears slightly out of trend: this could mean that this sample is not completely in the dirty limit or that its level of disorder has been underestimated.

**Discussion and Conclusions**

The drive to higher magnetic field superconducting $Nb_3Sn$ magnets for important applications in physics, chemistry and medical devices means that a better understanding of the high-field limitations of doped wires is highly desirable. Our recent studies showed that even the highest $J_c$ $Nb_3Sn$ strands still suffer from composition gradients that lead to property gradients[9]. For the HiLumi LHC upgrade, $Nb_3Sn$ wires were optimized for 12 T use by not pushing the heat treatment to maximum extent so as to prevent Sn leakage out of the sub-bundles into the surrounding high purity Cu. This involves some compromise to the irreversibility field and $H_{c2}$ but also has the benefit of maintaining a high grain boundary density essential for strong vortex pinning. However, higher field applications such as the proposed FCC require optimization of $J_c$ for fields closer to $H_{c2}$ (for the FCC the $Nb_3Sn$ target is 16 T) thus shifting the balance towards increasing $H_{c2}$ [23] and making the A15 layer more uniform. Until now it has been thought that the route to this goal requires maximum Sn stoichiometry. However, a full understanding of how Ta and/or Ti dopants act to enhance $H_{c2}$ was not clear and several contradictory hypotheses remained in play. To clarify the dopant site occupancy, we performed EXAFS characterizations of the highest performance $Nb_3Sn$ strands available that surprisingly revealed that Ta atoms not only substitute on the Nb sites but also occupy a significant fraction (~21-32%) of Sn sites in the $Nb_3Sn$, whereas Ti sits only on the Nb site. These findings raise the question whether or not the ability of Ta to occupy either site makes it compete with Sn for the Sn sites, slowing down the Sn diffusion that drives A15 phase conversion and producing Sn gradients at both layer and filament levels as observed in ref. 24. On the other hand, the strong preference of Ti for the Nb sites might favorably contribute to the Sn diffusion on its own sites, helping produce a more homogeneous phase with tighter distribution of properties[9]. These new insights suggest further studies to understand if and to what extent the choice of heat treatment temperature influences not only the distribution of the Sn but also the dopants distribution. Because of the evident off-stoichiometry of these strands, doping cannot be the only source of disorder in the A15 phase and antisite disorder seems to contribute significantly in increasing $H_{c2}$. In particular, the antisite disorder induced by Ti appears more effective than Ta in driving up $H_{c2}$. Interestingly, despite the lower 12 T performance, the double-doped (Ta+Ti) sample showed higher Kramer field ($H_k$) and upper critical field ($H_{c2}$) extrapolations with respect to similar single doped strands. This suggests that double-doped strands should be re-explored for 16-20 T applications, like FCC, re-investigating both the absolute dopant level and the optimal heat treatment.

These considerations and the markedly sub-stoichiometric Sn composition of the best performing Ti-doped internal-Sn strands (layer $J_c$ ~5.2 kA/mm$^2$, 4.2 K, 12 T) indicate that further investigation of global and local properties are needed in what will probably always be compositionally inhomogeneous wires. The striking possibility suggested by this site occupancy study, coupled to the compositional and high field specific heat study, is that antisite disorder in Ti-doped wires mitigates degradation of $H_{c2}$ in the Sn-deficient regions of the wires. We conclude that improvements of $J_c$ properties in the >16 T range at 4.2 K are still possible, also perhaps by using longer and higher HT temperatures to make the layers more compositionally uniform (recognizing of course that more effective diffusion barriers must be provided to protect the outer stabilizing Cu too). Although high field $J_c$ can also be enhanced by increasing the density and effectiveness of flux pinning sites by decreasing the grain size, adding pinning centers[25,26] or introducing point pinning by irradiation,[27,28] these techniques do not raise $H_{c2}$. Thus, consideration of these surprising site disorder results should also be incorporated into any enhanced pinning center approach in order to raise the high field $J_c$.

**Methods**

**Nb$_3$Sn bulk sample synthesis**. Constituent powders were first densified using a cold isostatic press (CIP) at 4 kpsi and then subjected to two separate heat treatments (HT) in a hot isostatic press (HIP) at 29 kpsi. The first was a two-step HT (650°C for 16 h. + 1200°C for 72 h). The second HT occurred at 1800°C for 24 h. The composition was verified by EDS (Energy Dispersive Spectroscopy) and confirmed to be between 24.3 and 25.3 at.% Sn.

**EXAFS set-up.** The EXAFS measurements of the wires using the 20-ID-B microprobe station were done using a Si (111) monochromator with energy resolution ΔE/E of 1.4x10$^{-4}$ in fluorescence mode using a 4-element vortex detector for the Ti K edge, and a bent Laue analyzer for the Ta L$_3$ edge. Since the filaments are surrounded by Cu, the bent Laue analyzer [29] was necessary to distinguish the Ta L$_{\alpha 1}$ line at 8146 eV from the Cu K$_{\alpha 1}$ line at 8046 eV. The Nb$_3$Sn standard was measured at beamline 20-BM also using a Si(111) monochromator. A focused beam of 0.5 mm$^2$ was provided by a toroidal mirror. The measurements were made in fluorescence using a 13-element Ge detector. Since the sample was thick, the results were corrected for self-absorption using the fluo algorithm [30] in the Athena analysis software[19]. Because of this correction, the $S_0^2$ values in Table I may be different from those needed to fit data for pure Nb$_3$Sn measured in transmission mode.

**Acknowledgements**

Sector 20 facilities at the Advanced Photon Source, and research at these facilities, are supported by the US Department of Energy - Basic Energy Sciences, the Canadian Light Source and its funding partners, and the Advanced Photon Source. Use of the Advanced Photon Source, an Office of Science User Facility operated for the U.S. Department of Energy (DOE) Office of Science by Argonne National Laboratory, was supported by the U.S. DOE under Contract No. DE-AC02-06CH11357. This material is based upon work partially supported by the U.S. Department of Energy, Office of Science, Office of High Energy Physics under Award Number DE-SC0012083. A portion of this work was performed at the National High Magnetic Field Laboratory, which is supported by National Science Foundation Cooperative Agreement No. DMR-1157490 and the State of Florida.

**Author Contributions**

S.M.H. performed EXAFS measurements and analysis. C.T. performed specific heat characterizations and disorder analysis. P.J.L. performed FESEM characterizations. Z-H.S. performed the EDS chemical analysis. M.B. synthetized and performed EDS analysis on the binary sample. A.K.G. reacted the sample and measured the transport properties. C.T., P.J.L. and D.C.L. designed the study. C.T., S.M.H., P.J.L. and D.C.L. prepared the manuscript. All authors discussed the results and implications and commented on the manuscript.

**Additional information**

Competing financial interests: The authors declare no competing financial interests.

**Figure Legends**

**Figure 1. Fourier transformed $k^2$ weighted EXAFS spectra from the Sn and Nb sites in pure $Nb_3Sn$ and sketch of the $Nb_3Sn$ structure with the main scattering paths**. The blue points and line in the spectra are the data, whereas the green line is a fit to the nearest neighbors using the $Nb_3Sn$ structure (the data are not phase corrected). The fitting window was between 1.7 and 3.3 Å for the Nb spectrum and from 1.9 to 3 Å for the Sn spectrum. We also show the $Nb_3Sn$ structure schematic showing the three nearby coordination shells for Nb (blue) and a single nearby shell of neighbors for Sn (orange).

**Figure 2. FESEM-BSE images and x-ray fluorescence maps of internal-Sn A15 wires.** FESEM-BSE image of a 54 sub-element strand overview (top left) and individual sub-elements of the Ti#1, Ti+Ta and Ta#1 doped strands. Bottom: x-ray fluorescence maps from a similar region in sample Ti#2 for Ti, and Sn (the apparent high Ti (Sn) composition on the right side of the sub-element is an artifact due to the experiment geometry).

**Figure 3. $T_c$-distributions at 0 and 15 T of the A15 phase in differently doped internal-Sn A15 wires.** The $T_c$-distributions were obtained analyzing specific heat characterizations of Ti#1, Ti+Ta and Ta#1 doped strands.

**Figure 4. Fourier transforms of the $k^2$ weighted $\chi(k)$ data for the Ta L3 edge using the $k$ range 2-11.5**. The blue points and line are the data, the green curve is a two-site fit as described in the text, and the red dashed curve is a fit to the Nb site only (fitting window from 1.9 to 3.5 Å; the data are not phase corrected).

**Figure 5. Fourier transforms of the $k^2$ weighted $\chi(k)$ data for the Ti K edge using the $k$ range 2-12.9**. The blue points and line are the data, and the green curve is the fit result for the Nb site (fitting window from 1.8 to 3.3 Å; the data are not phase corrected).

**Figure 6. Comparison between $H_{c2}(0)$ and disorder-related parameter.** $H_{c2}(0)$ estimated by WHH extrapolation versus $T_c(1-\eta^2)$ where $T_c$ is the critical temperature and $\eta$ is a weithed average of the long-range order parameters.

**Table I**. Parameters for the nearest neighbors in Nb$_3$Sn for the Nb and Sn sites. N and R are from crystallography. The other parameters are a result of fitting the data in Figure 1. $S_0^2$ and E$_0$ are an overall amplitude and energy shift to match the theory to the data[14].

| Scattering Path | N | R (Å) | R (fit) | $\sigma^2$ (Å$^2$) | $S_0^2$ | E$_0$ (eV) |
|---|---|---|---|---|---|---|
| Nb – Nb1 | 2 | 2.644 | 2.645 | 0.0060 | 1.06 | -1.80 |
| Nb – Sn | 4 | 2.956 | 2.948 | 0.0044 | 1.06 | -1.80 |
| Nb – Nb2 | 8 | 3.238 | 3.241 | 0.0100 | 1.06 | -1.80 |
| Sn – Nb | 12 | 2.956 | 2.951 | 0.0049 | 0.96 | 1.02 |

**Table II.** Description and Properties of Wire Samples. The chemical composition was estimated by EDS in the central part of the A15 layer. $J_c$ and $H_k$ values are self-field corrected.

| Design Subs/Stack | Sample ID | Dopant | Billet ID | Final HT | Nb at% | Sn at% | Ta at% | Ti at% | $T_{c,Onset}$ K | Non-Cu $J_c$(12T,4.2K) A/mm$^2$ | A15 layer $J_c$(12T,4.2K) A/mm$^2$ | $\mu_0 H_k$(4.2K) T | WHH $\mu_0 H_{c2}$(0K) T |
|---|---|---|---|---|---|---|---|---|---|---|---|---|---|
| 54/61 | Ta#1 | 4 at%Ta | 8781 | 640°C/40h | 72.46 | 25.17 | 2.37 | | 18.5 | 2712 | 4860 | 22.66 | 27.30 |
| | Ta+Ti | 4 at%Ta +1 at%Ti | 9362-5 | 640°C/40h | 71.41 | 24.64 | 2.60 | 1.35 | 18.1 | 2622 | 4528 | 24.59 | 28.77 |
| | Ti#1 | 2 at%Ti | 9415-BE | 640°C/40h | 74.87 | 23.40 | | 1.73 | 18.2 | 2872 | 5065 | 23.75 | 28.35 |
| 108/127 | Ti#2 | 2 at%Ti | 14895FE | 662°C/48h | 75.59 | 23.10 | | 1.31 | 17.9 | 3207 | 5173 | 25.45 | 29.29 |
| | Ta#2 | 4 at%Ta | 12879 | 665°C/48h | 73.09 | 23.39 | 3.52 | | 18.4 | 3138 | 5130 | 23.84 | 27.49 |

**Table III.** Fitting results for the Ta L$_3$ edge data.

| Sample ID, fit type | Scattering Path | N (Fixed) | R (fit) | $\sigma^2$ (Å$^2$) | Site Occupancy |
|---|---|---|---|---|---|
| Ta#1, two-site fit | Ta – Nb1 | 2 | 2.68 | 0.0042 | 0.70 |
| | Ta – Sn | 4 | 2.96 | 0.0097 | 0.70 |
| | Ta – Nb2 | 8 | 3.24 | 0.0087 | 0.70 |
| | Ta – Nb | 12 | 2.96 | 0.0097 | 0.30 |
| Ta+Ti, two-site fit | Ta – Nb1 | 2 | 2.64 | 0.0042 | 0.68 |
| | Ta – Sn | 4 | 2.96 | 0.0095 | 0.68 |
| | Ta – Nb2 | 8 | 3.24 | 0.0084 | 0.68 |
| | Ta – Nb | 12 | 2.96 | 0.0095 | 0.32 |
| Ta#2, two-site fit | Ta – Nb1 | 2 | 2.67 | 0.0044 | 0.79 |
| | Ta – Sn | 4 | 2.95 | 0.0078 | 0.79 |
| | Ta – Nb2 | 8 | 3.24 | 0.0091 | 0.79 |
| | Ta – Nb | 12 | 2.95 | 0.0078 | 0.21 |

**Table IV.** Fitting results for the Ti K edge data.

| Sample ID, fit type | Scattering Path | N (Fixed) | R (fit) | $\sigma^2$ (Å$^2$) | Site Occupancy |
|---|---|---|---|---|---|
| Ti #1, two-site fit | Ti – Nb1 | 2 | 2.67 | 0.0096 | 1.0 |
| | Ti – Sn | 4 | 2.91 | 0.0066 | 1.0 |
| | Ti – Nb2 | 8 | 3.25 | 0.0164 | 1.0 |
| Ti #2, two-site fit | Ti – Nb1 | 2 | 2.66 | 0.0054 | 1.0 |
| | Ti – Sn | 4 | 2.93 | 0.0059 | 1.0 |
| | Ti – Nb2 | 8 | 3.236 | 0.0110 | 1.0 |
| Ta+Ti, two-site fit | Ti – Nb1 | 2 | 2.66 | 0.0079 | 1.0 |
| | Ti – Sn | 4 | 2.92 | 0.0062 | 1.0 |
| | Ti – Nb2 | 8 | 3.240 | 0.0147 | 1.0 |

**Table V.** Composition ratios as obtained by EDS and EDS+EXAFS, site occupancies in the $A_3B$ structure taking into account antisite disorder (and exchange antisite for Ta+Ti) and long range order parameters.

| Design Subs/Stack | Sample ID | A15 Nb/Sn | (Nb+Ti+(1-x)Ta)/(Sn+xTa) | x | $c_{Nb}(A)$ | $c_{Ti}(A)$ | $c_{Ta}(A)$ | $c_{Sn}(A)$ | $c_{Sn}(B)$ | $c_{Ta}(B)$ | $c_{Nb}(B)$ | $\eta_{Nb}$ | $\eta_{Sn}$ | $\eta$ |
|---|---|---|---|---|---|---|---|---|---|---|---|---|---|---|
| 54/61 | Ta#1 | 2.879 | 2.864 | 0.30 | 0.966 | | 0.022 | 0.012 | 0.972 | 0.028 | | 0.966 | 0.960 | 0.965 |
| | Ta+Ti | 2.898 | 2.926 | 0.32 | 0.942 [0.952] | 0.018 [0.018] | 0.024 [0.024] | 0.017 [0.006] | 0.936 [0.967] | 0.033 [0.033] | 0.031 [0] | 0.911 [0.952] | 0.919 [0.960] | 0.913 |
| | Ti#1 | 3.200 | 3.336 | | 0.977 | 0.023 | | | 0.936 | | 0.064 | 0.913 | 0.936 | 0.919 |
| 108/127 | Ti#2 | 3.272 | 3.414 | | 0.983 | 0.017 | | | 0.924 | | 0.076 | 0.907 | 0.924 | 0.911 |
| | Ta#2 | 3.125 | 3.144 | 0.21 | 0.963 | | 0.037 | | 0.936 | 0.030 | 0.035 | 0.928 | 0.936 | 0.930 |

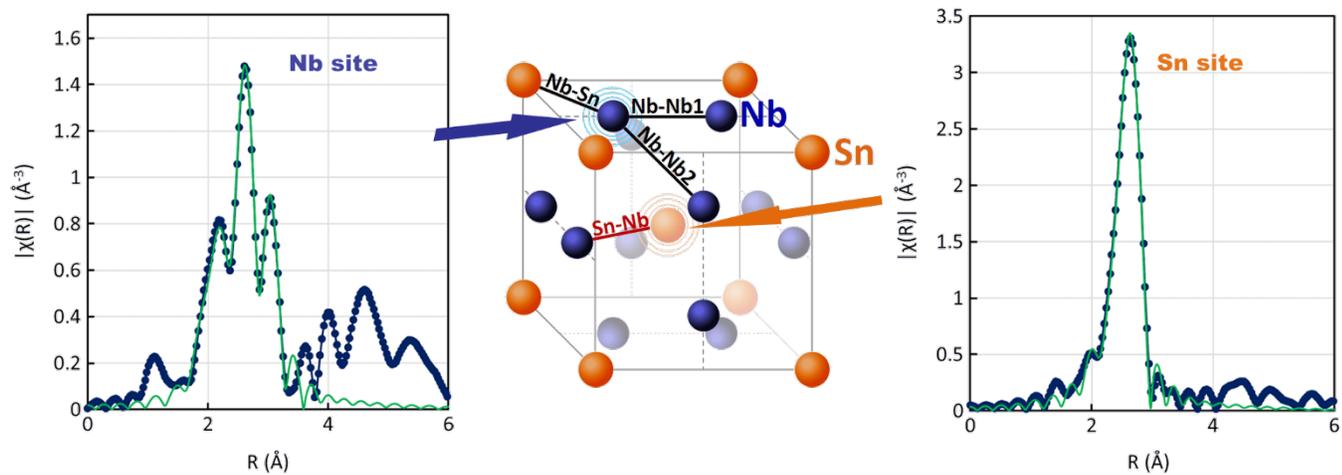

**Figure 1**

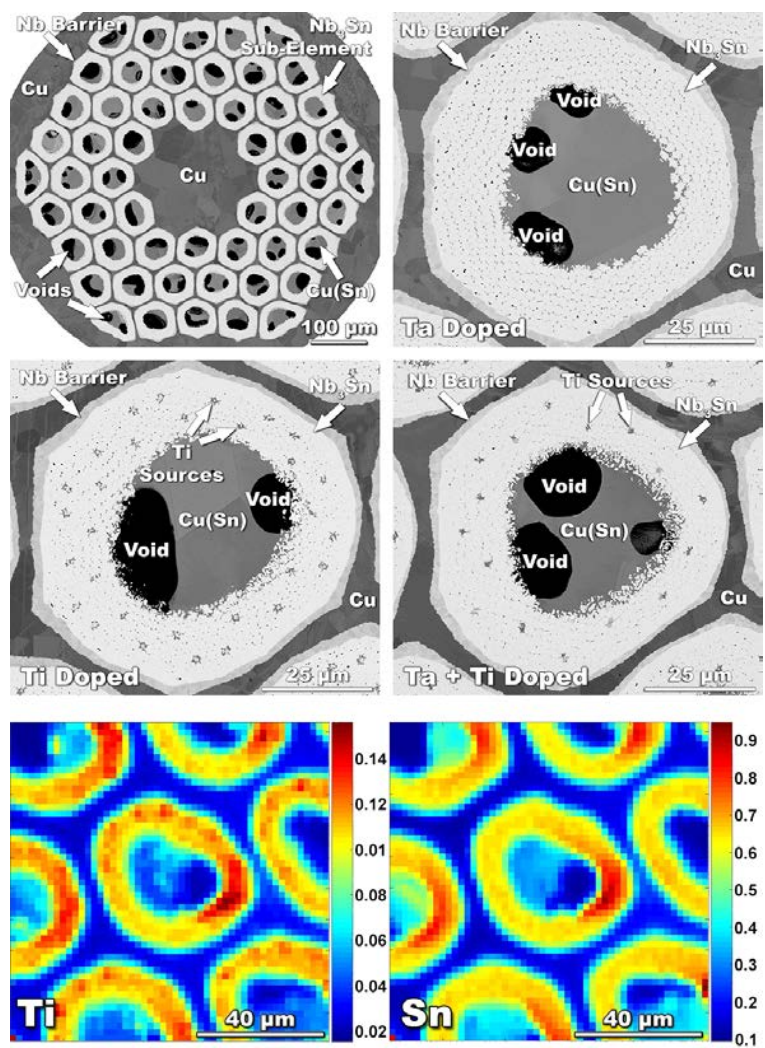

**Figure 2**

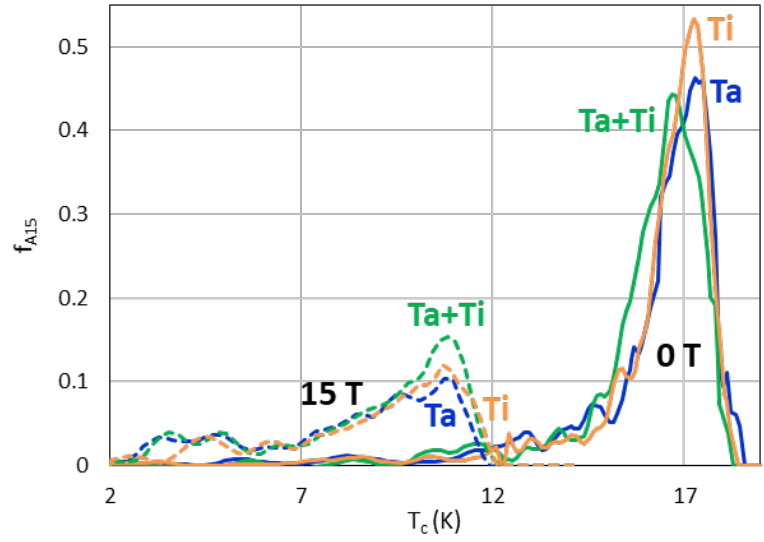

**Figure 3**

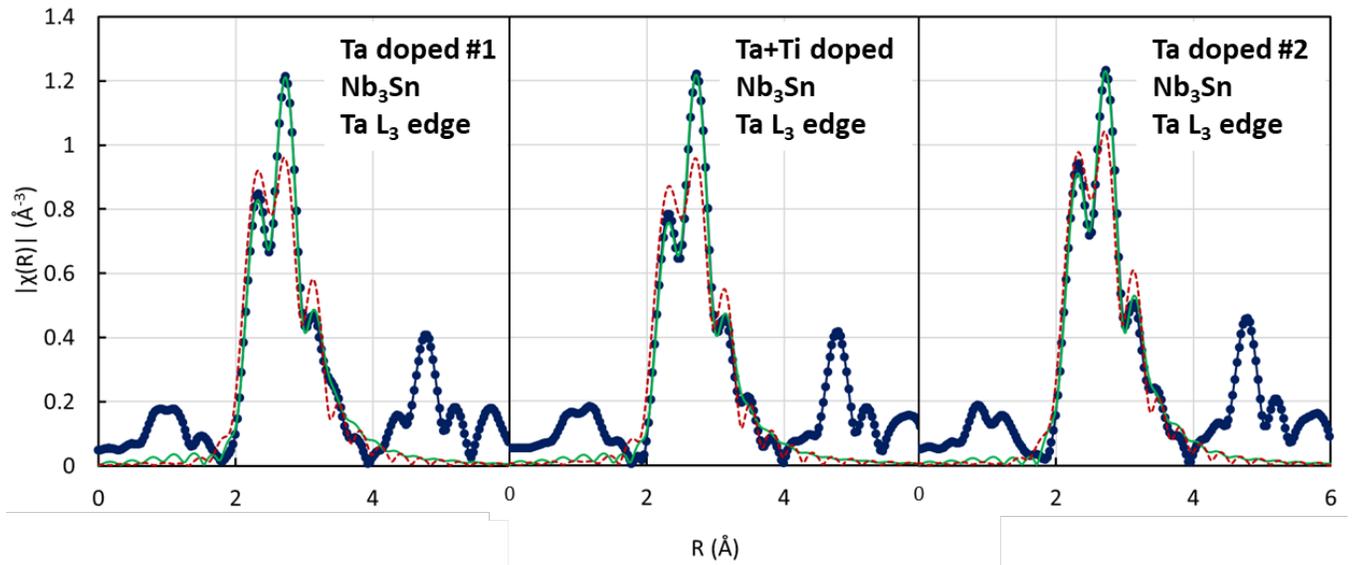

**Figure 4**

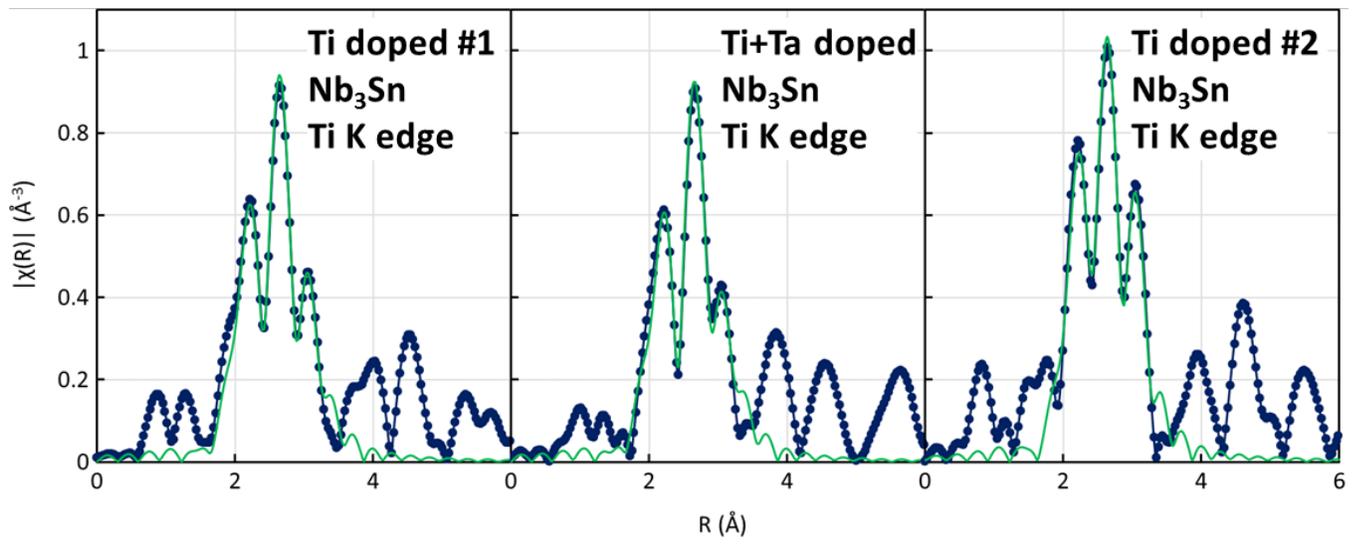

**Figure 5**

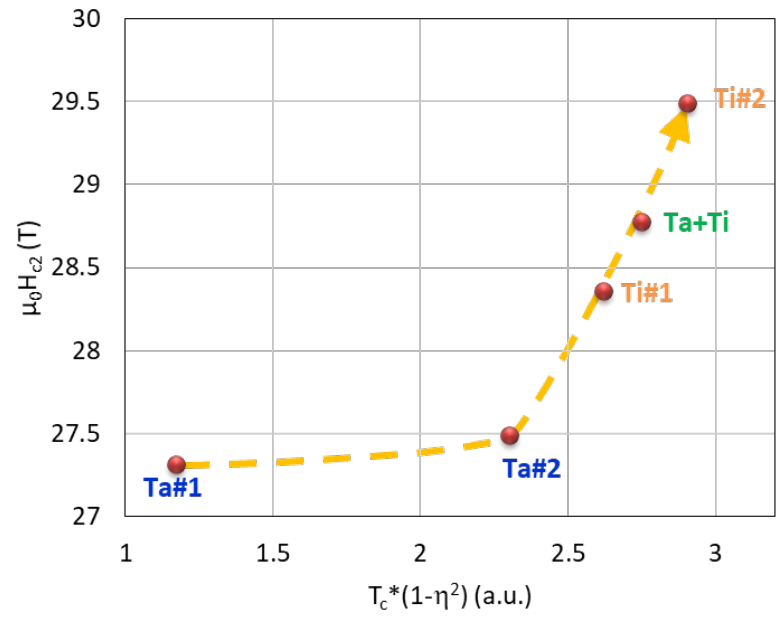

**Figure 6**